\documentclass[11pt]{article}
\usepackage{setspace}
\usepackage{amssymb}
\usepackage{amsmath}
\usepackage{amsfonts}

\def\be{\begin{equation}}

\def\ee{\end{equation}}

\def\dd{\partial}

\newcommand\of[1]{\left( #1 \right)}

\def\bea{\begin{eqnarray}}

\def\eea{\end{eqnarray}}

\newcommand\T{ {\hbox{\tiny{T}}}}
\newcommand\TT{{\hbox{\tiny{TT}}}}
\newcommand\LL{  {\hbox{\tiny{L}}}}
\begin{document}

\singlespace

\begin{flushright} BRX TH-680 \\
CALT-TH 2014-147
\end{flushright}

\vspace*{.3in}

\begin{center}

{\Large\bf Unconstrained canonical action for, and positive energy of, massive spin 2 }

{\large S.\ Deser}

{\it 
Walter Burke Institute for Theoretical Physics, \\
California Institute of Technology, Pasadena, CA 91125; \\
Physics Department,  Brandeis University, Waltham, MA 02454 \\
{\tt deser@brandeis.edu}
}

\end{center}

\begin{abstract}
Filling a much-needed gap, we exhibit the $D=4$ Fierz-Pauli (FP) massive $s=2$ action, and its -- manifestly positive -- energy, in terms of its $2s+1=5$ unconstrained helicity ($2$,$1$,$0$) excitations, after reducing and diagonalizing the troublesome helicity-0 sector.
\end{abstract}

\section{Introduction}

We begin with an apologia, since there is nothing still not known about the FP model [1].  However, there seems to be no published derivation\footnote{This was actually done [2] for the more general case of the system embedded in dS, rather than flat, space; however that derivation involved (an even number of) steps using inverse powers of $\Lambda$, hence singular and not applicable here; its final result of course does limit to ours.  It was also shown in detail here that the helicity $0$ mode can be removed by suitably tuning $m^2/\Lambda$ in dS.} from its covariant and highly constrained form to the final unconstrained
canonical action in terms of its $2s+1=5$ helicity $(\pm 2,\pm 1,0)$ components. That form will displays that each mode propagates correctly and has manifestly positive energy, in the usual $L=p \, \dot q - H$, $H= \frac{1}{2}\, [p^2+ q\, (-\nabla^2+m^2)\, q]$ form. While [1] realized that positive energy was essential, it was displayed rather opaquely; a subsequent formulation [3] was likewise less than transparently presented (and contains distracting typos). lndeed, proper use of the constraints is not altogether trivial, making the correct process instructive as a (minor) exercise in (free) field theory.

Historically, it was not until FP's 1939 work that the problem of representing massive spins $>1$ by tensor fields -- involving many more than $2s+1$ components -- was raised, let alone solved. Given the current interest in massive gravity (mGR) with Einstein kinetic terms plus non-derivative mass terms involving a fixed, say flat, background, our summary may be useful. lndeed, this is a good place to note that-contrary to statements in the mGR literature -- the mass terms destroy the whole (ADM) asymptotic energy formulation of GR as a $2D$ surface integral at spatial infinity, just as the Coulomb asymptotic integral that counts total charge is lost in massive (Proca) vector theory: the Newtonian/Coulomb fields decay much too fast there for these integrals to contribute at all.

\section{The Derivation}

The action and field equations of the theory are the sum of linearized GR and the -unique-mass term that eliminates the 6th, ghost helicity $0$, degree of freedom (DoF). We work throughout in 1st order, $3+1$, canonical form, which simplifies the procedure and indeed starts directly in terms of the 6 conjugate pairs $(\pi^{ij}, h_{ij})$ rather than the 10 covariant $h_{\mu\nu}$. The action is (see e.g., [2])
\begin{equation}
\begin{aligned}
I &= \int d^4x\,  \left\{ \pi^{ij} \dot h_{ij} - H(\pi,h)\right\}, \\
H &= \ ^3R_Q + \of{\pi_{ij}^2 - \frac{1}{2}\, \pi^2} + 4 \, n\,  R_0 + 2\, N_i \, \dd_j \pi^{ij} + \frac{1}{4} \, m^2 \, (h_{ij}^2 - h_{ii}^2 - 4 \, n\, h_{ii} - 2 N_i^2),
\end{aligned}
\end{equation}
where (under the integral)
\begin{equation}
\ ^3 R_Q = \frac{1}{2} \, h_{ij} \, G^L_{ij}  = -\frac{1}{4}\,  [h_{\TT} \, \nabla^2 h_{\TT} - h_\T \, \nabla^2 h_\T],  \, \, \, \, \, \, \, \, \, \, R_0 = (m^2 - \nabla^2) \, h_\T + m^2\,  h_\LL;
\end{equation}
$n \sim \frac{1}{2} \, h_{00}$ is a Lagrange multiplier enforcing the linear constraint $R_0 =0$ , while $N_i=h_{0i}$ becomes an auxiliary field to be eliminated by completing squares, leaving only the six $(\pi,h)$ pairs -- indeed our whole process consists of juggling quadratic forms. Finally, we recall that the linearized 3D Einstein tensor $G^L_{ij} (h_{lm})$ is both identically conserved and independent of the longitudinal, gauge, parts of $h_{lm}$; The second ingredient, essential to the separation of the various helicity DoF in (1), is the usual orthogonal decomposition of any symmetric 3-tensor,
\begin{equation}
S_{ij} = S_{ij}^{\TT} + \frac{1}{2}\, (\delta_{ij} - \hat \dd_i \hat \dd_j) S^\T + [ \hat \dd_j S^\T_i + \hat \dd_j S^\T_{i}] 
+ \hat \dd_i \hat \dd_j S^\LL,  \, \, \, \, \, \dd_i \, S^\T_i \equiv 0, \, \, \, \, \, \hat \dd_i \equiv \dd_i/\sqrt{\nabla^2}. 
\end{equation}

Completing squares in (1) removes the $N_i$ dependence of $H$ in favor of adding the term $2m^{-2} (\dd_j \pi^{ij})^2$ to $H$. There remains the elimination of the $R_0$ constraint, hence of one linear combination of the two helicity $0$ ($T$, $L$) modes. lt will be equally essential to use $\dot R_0 =0$ to further
eliminate one combination of their conjugate momenta $(\pi^\T,\pi^\LL)$ using $ \dot h \sim \pi$; constraints ``strike twice" in our 1st order form, since they are valid for all times\footnote{
That the original number, $6$, of $\pi^{ij} \dot h_{ij}$ kinetic terms decreases by one for every constraint is just Darboux's theorem on quadratic forms; in massless theory there are $4$ constraints, leaving just the two $(\pi^{\TT}, q^{\TT})$ pairs. We will see this more explicitly below.}.

The task before us then is to decompose (1) into a sum of three -- non- interacting -- orthogonal, two DoF sectors: Helicity $\pm 2$ (TT), helicity $\pm 1$ ``$T_i$", and the $(T,L)$ helicity-$0$. The latter's Hamiltonian is the source of difficulty, being a priori non-positive before using the $R_0(T,L)=0$ constraint. To keep the discussion compact, we first dispose of the helicity $>0$ sectors: that of TT is trivial to obtain, being unconstrained; we simply add up the TT terms in (1); dropping ``TT", we have
\begin{equation}
L = \pi^{ij} \,  \dot h_{ij} - H \, \, \, \, \, \, \, \, \, \, H = \pi_{ij}^2 + \frac{1}{4} \, \of{ h_{ij,k}^2 + m^2 \, h_{ij}^2} \ge 0;
\end{equation}

Note that $H$ vanishes only for TT vacuum, $\pi=0=h$. The same is true of the transverse vector ($T_i$) part, though it still requires some field redefinitions to achieve the same final form; here (1) also easily yields (omitting ``$T_i$")
\begin{equation}
H = \pi^2 + 2 \, m^{-2} \, \of{\pi^{ij}_{\, \, \, ,j}}^2 + \frac{1}{2} \, m^2 \, h^2 \ge 0;
\end{equation}
while not (yet) very pretty, this H is also positive and vanishes at ($T_i$) vacuum, a result unaffected by the further field redefinitions required to reach the final $p\, \dot q - H$ form; we outline the process in the Appendix. [Recall, however, that correct energy functional form is only reached when the ``$(p,q)$" variables are redefined to ensure that the associated kinetic, ``$p\, \dot q$", term is itself free of unwanted numerical coefficients.]

We now face the final, $H(T,L)$, sector, where $R_0$ must be used -- twice. There,
\begin{equation}
\begin{aligned}
I[T,L] &= \int d^4 x \left[\frac{1}{2} \, \pi^\T \,  \dot h_\T +  \pi^\LL \, \dot h_\LL - V(h_\T, h_\LL) - K(\pi^\T, \pi^\LL)\right],  \\
4 \, V(h) &\equiv h_\T \, \of{\nabla^2 - m^2} \, h_\T - 2 \, h_\T \, h_\LL ,  \, \, \, \, \, \, \, \, \, \,  K(\pi) \equiv \frac{1}{2}\, \of{\pi^{\LL}}^2 - 2 \, \pi^\T \, \pi^\LL + 2 \, \pi^\LL \, \of{-m^{-2}\, \nabla^2} \, \pi^\LL.
\end{aligned}
\end{equation}

We now show that both potential and kinetic parts of $H$ are positive, using the $R_0$ constraint and its time derivative respectively.
Eliminating $h_\LL$ yields
\begin{equation}
V(h_\LL,h_\T) = \frac{3}{8}  \, \left[ \of{h_{\T , i}}^2 + m^2 \, h_\T^2\right] \ge 0;
\end{equation}
again, $V$ only vanishes at vacuum, $h_\T=0$. 
 Next we find the $\dot R_0 = 0$ constraint between $\pi^\T$ and $\pi^\LL$:  The two field equations for $\pi \sim \dot  h$ obtained by varying (6) w.r.t. the $\pi$ are
\begin{equation}
\dot h_\LL - \pi^\LL + \pi^\T + 4 \, m^{-2} \, \nabla^2 \pi^\LL = 0, \, \, \, \, \, \, \, \, \, \, \dot h_\T + 2 \, \pi^\LL =0.
\end{equation}

Taking their appropriate vanishing linear combination, we learn that
\begin{equation}
m^2 \, \pi^T  = \of{-2 \, \nabla^2 - m^2}\, \pi^\LL;
\end{equation}
hence finally 
 \begin{equation}
K(\pi^\LL,\pi^\T) \rightarrow K(\pi^\LL) =\frac{3}{2} \, \of{\pi^L}^2 \ge 0. 
  \end{equation}
We have now established $E\ge 0$ for the full theory, but one task is still to be completed: putting the helicity action into exact
$p\, \dot q- H(p,q)$ form. Even before this is done, one can already see that the (second order) field equations are uniformly $(\Box-m^2)\, h=0$, but it is an amusing exercise -- as well as a check on the result -- to do so. Using (9), it is  easy to translate the $(T,L)$ sector's $\pi^\LL\, \dot h_\LL +\frac{1}{2} \, \pi^\T\,  \dot h_\T$ into $\pi^\LL\,  \dot h_\T$ form.  At this penultimate point,
\begin{equation}
L (T,L) = -\frac{3}{2} \, \pi^\LL \, \dot h_\T - \frac{3}{2} \, [ \of{\pi^\LL}^2 + \frac{1}{4}\,  \of{h_{T,i,i}}^2 + \frac{1}{4} \, m^2 \, h_\T^2 ]; 
 \end{equation}
the obvious rescaling $(\pi,h) \rightarrow \sqrt{2/3} \, (-\pi,h)$ achieves the desired final canonical form of the helicity $0$ sector,
\begin{equation}
L(0) = \pi \, \dot h - H(\pi,h), \, \, \, \, \, \, \, \, \, \, H \equiv \pi^2 + \frac{1}{4} \, h \, \of{-\nabla^2 + m^2} \, h.
\end{equation}
Together with the vector mode in the Appendix, then, the total action is
\begin{equation}
L = \sum_{A=1}^5 p^A \, \dot q_A - \frac{1}{2}\, [ (p^A)^2 + q_A (-\nabla^2 + m^2) \, q_A ],
\end{equation}
after the (cosmetic) rescaling $\pi \rightarrow p^A/\sqrt{2}$, $h\rightarrow q_A \, \sqrt{2}$.

\section{Summary}
The physical correctness of the massive $s=2$ FP model has been displayed: each of its $2s+1=5$ helicity excitations obey $(\Box-m^2)\, h=0$, $E \ge 0$.

\section{Appendix: The helicity $\pm 1$ sector}

We consider here the remaining, helicity $1$, subspace involving only the ``$T_i$" parts of $(\pi,h)$ in (1). Clearly, neither $h\, G_\LL (h)$  nor the $R_0$ constraint involve $h_{\T\, i}$; only the mass term does: it contains $\frac{1}{2} \, m^2 \, (h_{\T \, i})^2$.  Its $\pi$ sector involves $(\pi^\T_{i,j})^2$ as well as the quadratic term $2\, N^\T_i \dd_j \, \pi^{ij}$. Using the $\frac{1}{2}\, m^2 (N^\T_i)^2$ from the mass term,  we complete the square to leave a net contribution $m^{-2}(\pi^{ij}_{\, \, \,  ,j})^2 \sim 2 \, m^{-2} \, (\pi^\T_i \, \nabla^2 \pi^\T_i)$ there. At this point, then, dropping the $T_i$ indices, we find
\begin{equation}
L(T_i)= -2 \, \pi \, \dot h - \frac{1}{2} \, [m^2 h^2 + 4 \, m^{-2} \, \pi \, \of{m^2 - \nabla^2} \pi]  ;
\end{equation}
the necessary redefinition is obvious: 
\begin{equation}
\pi \rightarrow -\frac{1}{2} \, m \of{m^2 -\nabla^2}^{-1/2} \, \pi  \, \, \, \, \, \, \, \, \, \, h \rightarrow m^{-1} \, \sqrt{m^2 - \nabla^2} \, h
\end{equation}
leads to the desired helicity $1$ canonical Lagrangian,
\begin{equation}
L(T_i) \rightarrow \, \pi \, \dot h - \frac{1}{2} \, [ \pi^2 + h \, \of{m^2 - \nabla^2} \, h ].
\end{equation}

\subsection*{Acknowledgements}
This work was supported in part by Grants NSF PHY- 1266107 and DOE \# DE-SC0011632. Collaboration with A. Waldron on [2] facilitated the present effort, as did composing help from J. Franklin and CPU help from G. Conrad.

\end{document}